\documentclass[onecolumn, pdftex]{revtex4}

\usepackage[pdftex]{graphicx}
\usepackage{amsmath}
\usepackage{hyperref}
\usepackage{doi}

\begin{document}
\title{Mass composition of cosmic rays determined by muon fraction with $\varepsilon_{thr.} \geq$ 1 GeV in air showers with energy greater than 5 EeV}

\author{S. Knurenko}
\author{I. Petrov$^{*}$}

\affiliation{Yu.G. Shafer Institute of Cosmophysical Research and Aeronomy SB RAS, 31 Lenin ave. 677980 Yakutsk, Russia}

\email{igor.petrov@ikfia.ysn.ru}

\begin{abstract}
	The ratio of the number of muons with a threshold of 1 GeV and charged particles at a distance of 600 m from the axis is analyzed. Air showers with energies above 5 EeV and zenith angles with less than 60 degrees are considered. Comparison of experimental data with calculations by the QGSJETII-04 model for various primary nuclei, including a gamma ray, showed that the mass composition of cosmic rays up to energies of $\sim$ 10 EeV mainly consists of protons and helium nuclei, with a small number of heavy nuclei. Data include air showers with a very low muon content, which are assumed to be produced by primary gamma rays.
\end{abstract}

\maketitle

\section{Introduction}

Astrophysics of cosmic rays of highest energies above 1 EeV is an important area of research for gaining knowledge about evolution and processes occurring in the Universe \cite{Ginzburg1970192,Ginzburg199942,Gelmini2007173,Bhattacharjee2000109,Halzen20021025}.

Recent observations made by telescopes find interesting structures --- some bubbles \cite{LiApJ2019873}, which are possibly sources of intergalactic cosmic rays with energies greater than 5 EeV along with other active astronomical objects: supernova remnants, compact jets of active galactic nuclei \cite{Mannheim200163}, clusters of galaxies \cite{Kang1996456}, radio galaxies \cite{Rachen1993272}, and gamma ray bursts \cite{Waxman199575}. In the past it has been assumed that protons dominate in this region \cite{Berezinsky196928}, while at energies less than 0.1 EeV, galactic cosmic rays consist of a mix of nuclei \cite{Stanev1993274}.

At the same time, the astrophysical aspect of cosmic rays of highest energies is still a little studied area. There is no reliable information about the nature of the origin of cosmic rays of highest energies. In particular, sources and their location in the Universe are not known well enough. The mechanism of generation, acceleration and propagation of particles has not been reliably established. It is unknown how particles interact during their propagation in outer space with magnetic fields and cosmic microwave background \cite{Berezinsky196928, Becker2008458, Ahlers201034, Kotera201010, Gelmini2008106}.

At the largest experiments for the study of air showers, recently, interesting data have been obtained on cosmic rays in the energy region above 1 EeV: the cosmic ray spectrum \cite{Aab2017038, Abbasi201680}, the mass composition \cite{Knurenko201964, Hanlon201819, Berezhko201231, Kampert201235, Knurenko201599, Thoudam2016595,Aab201490,Abbasi201564}, and the anisotropy of the arrival of particles with highest energies \cite{Knurenko2009196, Ivanov200887, Abreu2012203} . Verification of the obtained data is still small and further research in these areas is required.

The following results have been obtained in the energy range 1-10 EeV: according to the Pierre Auger Observatory --- mass composition is mainly protons \cite{Aab201490}; according to the Telescope Array experiment --- mass composition is a mixture of protons and helium nuclei \cite{Abbasi201564}. The flux of cosmic rays according to the Yakutsk experiment data: in the energy range 10-100 PeV --- a mix of light and heavy nuclei; in the energy range 0.1-10 EeV --- a mix of protons and helium nuclei \cite{Knurenko201964, Berezhko201231}. 

The results obtained indirectly largely depend on the equipment and methods of the experiment, the atmospheric conditions, methods for processing the experimental data, hadronic interaction models, and other factors. Therefore, to verify the results obtained earlier, it is important to obtain mass composition using a different technique and a different component of air shower, e.g. muons. Muons are measured at the Yakutsk array almost the entire year. The technique does not depend on weather conditions, as in the case of Cherenkov light at the Yakutsk array or fluorescence light at the Auger and TA arrays. The total time of optical measurements, for comparison, at the Yakutsk is only $\sim$ 10$\%$ of the total time of charged particles measurements.

It is known that the muon component is sensitive to the mass composition of the primary particles that produces air shower \cite{Knurenko2018107}, as was shown by calculations using the QGSJETII-04 model \cite{Ostapchenko201183} for the primary proton and iron nucleus. First of all, this refers to the relative content of muons, i.e. the ratio of the muon flux density to the charged shower component at a distance of 600 m from the shower axis $\rho_{\mu}/\rho_{\mu+e}$. This parameter is used at the Yakutsk experiment to analyze the mass composition of cosmic rays of highest energies. According to calculations, a joint analysis of the muon content with the longitudinal development of air shower at fixed energy and in the case of vertical shower can provide a reliable estimate of the mass composition of cosmic rays. It even can separate the primary particles by atomic weight \cite{Atrashkevich198133}, i.e. to distinguish air showers produced by gamma rays, protons, and iron nuclei \cite{Knurenko2009196}. We performed such analysis for the Yakutsk array data and estimated Cherenkov light, total charged and muon components \cite{Knurenko2018107}. The data for this work consist of air showers with energies above 5 EeV and zenith angles $\theta \leq 60^{\circ}$. In the region of these energies, “dip” and “bump” type irregularities are formed in the spectrum of cosmic rays \cite{Aab2017038,Abbasi201680}, which has been interpreted as the interaction of galactic protons with the cosmic microwave background of the Universe \cite{Rachen1993272, Berezinsky200421}. This hypothesis can be confirmed only by having information about the mass composition of cosmic rays in the energy range 5-50 EeV. In this work, for this purpose, an analysis of the relative muon content in showers of highest energies was carried out.

\section{Yakutsk array for air shower registration}

\begin{figure}
	\includegraphics[width=0.7\textwidth]{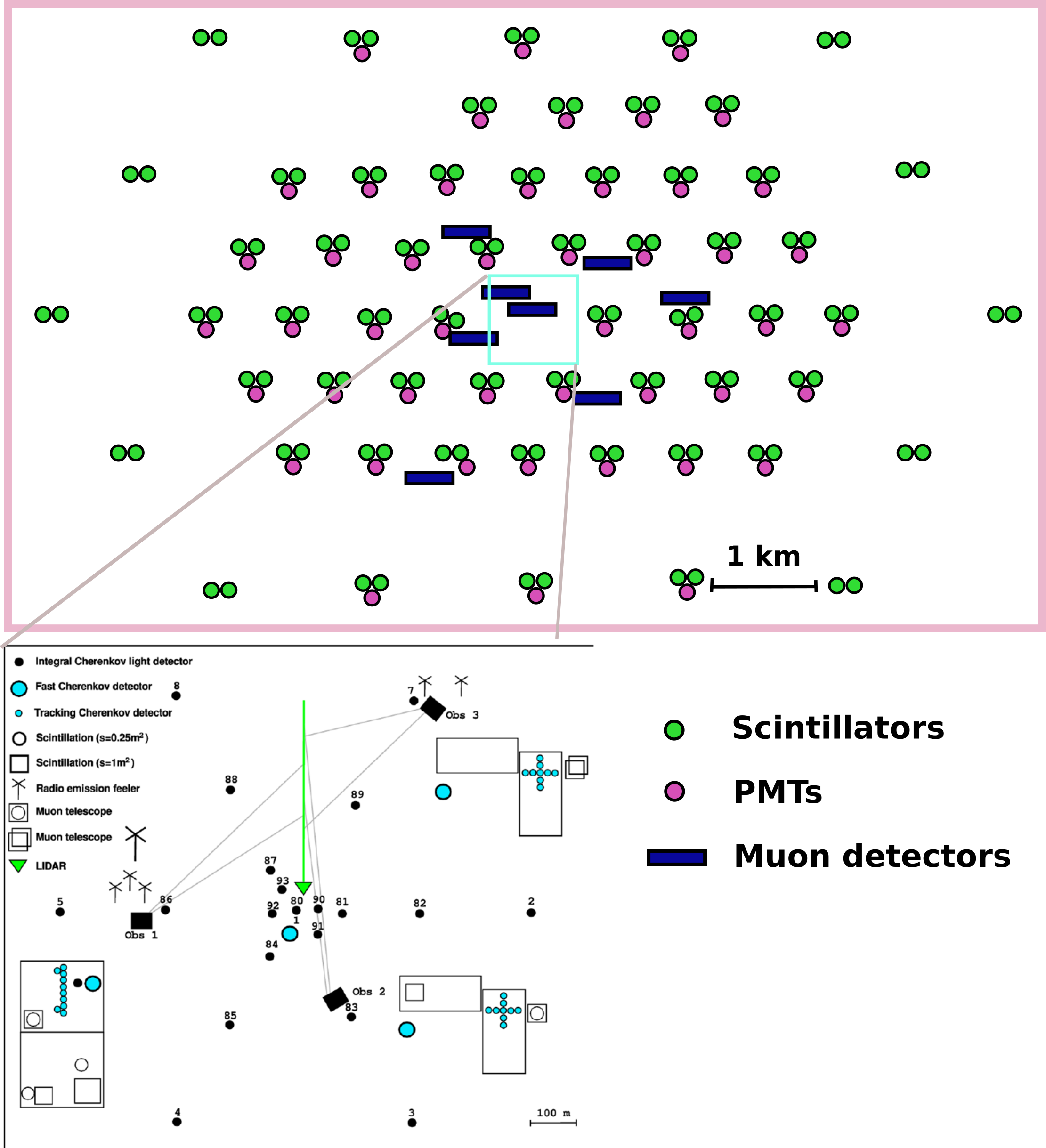}
	\caption{The layout of the observation stations at the Yakutsk array. The double circles show the stations at the array, two scintillation detectors are located at each station, and the stations with Cherenkov detectors are shown with triple circles. Rectangles show muon detectors. In the center of the array is the Small Cherenkov array (shown by a blue square). Small Cherenkov array consist of muon detectors, Cherenkov light detectors, and scintillation detectors with 50, 100, and 250 m spacing. \label{ykt-mc-experiment}}
\end{figure}

Fig. \ref{ykt-mc-experiment} shows the location of observation stations at the Yakutsk array. The array consists of 120 scintillation detectors with a threshold of 10 MeV and a receiving area of 2 m$^{2}$. Each station has two scintillation detectors and one Cherenkov detector with a photocathode receiving area of 176 cm$ ^{2} $ or 530 cm$ ^{2} $ \cite{Artamonov199412}.

Subfigure (Fig. \ref{ykt-mc-experiment}) shows the central part of the array in close-up. Over an area of $\sim$ 0.8 km$ ^{2} $, registration station for charged particles, air shower Cherenkov light (integrated and differential detectors), muon telescopes, cameras obscura and radio antennas are located \cite{Egorov2018301, Knurenko2017230}. The distances between the stations are 50, 100, 250 and 500 m. The central area is filled with a small Cherenkov array with its own independent trigger that detects air shower events in the energy range from 2 PeV to 1 EeV using the trigger from the detectors of Cherenkov light \cite{Knurenko199846}.

The Yakutsk experiment in various configurations has been operating since 1970, continuously registering ultra-high energy air shower events. In the 70s of the last century, registering air showers on an area of 3.5 km$ ^{2} $. From 1974 to 1994 --- over an area of $\sim$ 20 km$ ^{2} $. From 1994 to 2004, registration was carried out over an area of $\sim$ 18 km$ ^{2} $, and after 2004, the area decreased to the current $\sim$ 13 km$ ^{2} $. The Yakutsk experiment differs from other large experiments by its hybrid measurements: the charged and muon components, Cherenkov and radio emission from air showers. The system of observation stations selects shower events in ground-based scintillation detectors using a single “500” trigger (the trigger uses detectors with 500 m spacing) --- matching scheme C6 = C2 + C3. This scheme is a coincidence of two local triggers: C2 and C3. C2 --- coincidence of signals from two scintillation detectors for 4 $\mu s$ in the observation station; C3 --- coincidence of three or more stations within 40 $\mu s$ located in a triangle. Satisfying these two conditions is called the “C6” trigger. From the air shower events selected in this way, a primary database of the Yakutsk experiment was created. After a preliminary analysis of the data, a set of showers was used to find the parameters of individual air showers: arrival direction, coordinates of the core, power of the shower --- total number of charged particles $N_s$, and the classification parameter $\rho_{\mu+e}$ --- density of the flux of charged particles at distances of 600 m from the shower axis.

We used the lateral distribution function (LDF) of charged particles perpendicular to the axis of the shower to locate the core:
\begin{equation}\label{ykt_eq_1}
\rho_{\mu+e} (R) = 
\rho_{\mu+e}
\cdot \frac{600}
{r}
\cdot
\left(
\frac{1+\frac{600}{R_{0}}}
{1+\frac{r}{R_{0}}}
\right)^{b_{s}-1}
\end{equation}
where $R_{0}$ is the Moliere radius in meters, $b_{s}$ is the slope characteristic of the lateral development function (LDF) of charged particles, which depends on the classification parameter $\rho_{\mu+e}$ and the zenith angle $\theta$, according to formula (\ref{ykt_eq_2}):
\begin{equation}\label{ykt_eq_2}
b_{s} (\theta, \rho_{\mu+e}) = 
(1.38 \pm 0.06) 
+ (2.16\pm 0.17)
\cdot\cos{\theta}
+ (0.5 \pm 0.03)
\cdot \log_{10}{\rho_{\mu+e}(600)}
\end{equation}
This expression was obtained empirically from measurements of showers with different energies and different zenith angles \cite{Glushkov199357}.

In the table \ref{ykt_tab_1} as an example, a part of the data is shown --- 2009-2015. Table \ref{ykt_tab_1} shows the observation time, statistics of recorded air showers, the percentage of analyzed showers, the number of showers with registered muons and Cherenkov light, the observation time of the Cherenkov component and the number of showers with energies above 10 EeV. N --- total number of registered air shower events, $N_{muon}$ --- number of air shower events with registered muon component, and $N_{ch}$ --- number of air shower events with registered Cherenkov component. 

\begin{table}
	\caption{\label{ykt_tab_1} Air shower statistics registered at the Yakutsk array}
	\begin{tabular}{|c|c|c|c|c|c|c|c|}
		\hline 
		Year  &  Obs. time & N       & Data proc. & $N_{muon}$ & $N_{ch}$ & Cherenkov obs. time & Events E$\geq$10EeV \\ 
		\hline 
		09-10 & 6154    &  113138 & 87$\%$     & 60618      & 9897     & 622              & 10  \\ 
		\hline 
		10-11 & 6455    &  137830 & 89$\%$     & 56130      &     8611 & 508              & 15  \\ 
		\hline 
		11-12 & 6534    & 155351  & 91$\%$     & 54559      &  9227    & 482              & 15  \\ 
		\hline 
		12-13 & 6515    & 149381  & 92$\%$     & 89430      & 10219    & 592              & 17  \\ 
		\hline 
		13-14 & 6446    & 147589  & 91$\%$     & 72110      & 7164     & 396              & 15   \\ 
		\hline 
		14-15 & 6365    &  140101 & 72$\%$     & 82392      & 7838     & 429              & 15  \\ 
		\hline 
	\end{tabular} 
\end{table}

Currently, the Yakutsk array has an area of $\sim$ 13 km$ ^{2} $ and is medium in size, between compact arrays with an area of s $\leq$ 1 km$ ^{2} $ and giant arrays with s $\geq$ 1000 km$ ^{2} $ \cite{Abraham2004523, AbuZayyad2012689}. The array with such area is able to efficiently detect showers in energy range from 1 PeV to 100 EeV, and it allows us to study both galactic and metagalactic cosmic rays.

\section{Muon registration at the Yakutsk experiment}
\label{sec_muon}

The registration of muons with a threshold of $\varepsilon_{thr.} \geq$ 1 GeV at the Yakutsk experiment was started in 1974 by three observation points with an area of 16 m$ ^{2} $ each. Two points were located at a distance of 500 m and one at a distance of 300 m from the array center. Later, in 1994, three more stations were built, each with an area of 20 m$ ^{2} $. Two points were located at a distance of 800 m and one point at a distance of 500 m from the array center. In 1998, another station for registering muons with $\varepsilon_{thr.} \geq$ 0.5 GeV and an area of 190 m$ ^{2} $ was added to the existing stations. It is located at a distance of 150 m from the array center \cite{Artamonov199412, Ivanov201065}.

\begin{figure}
	\includegraphics[width=0.7\textwidth]{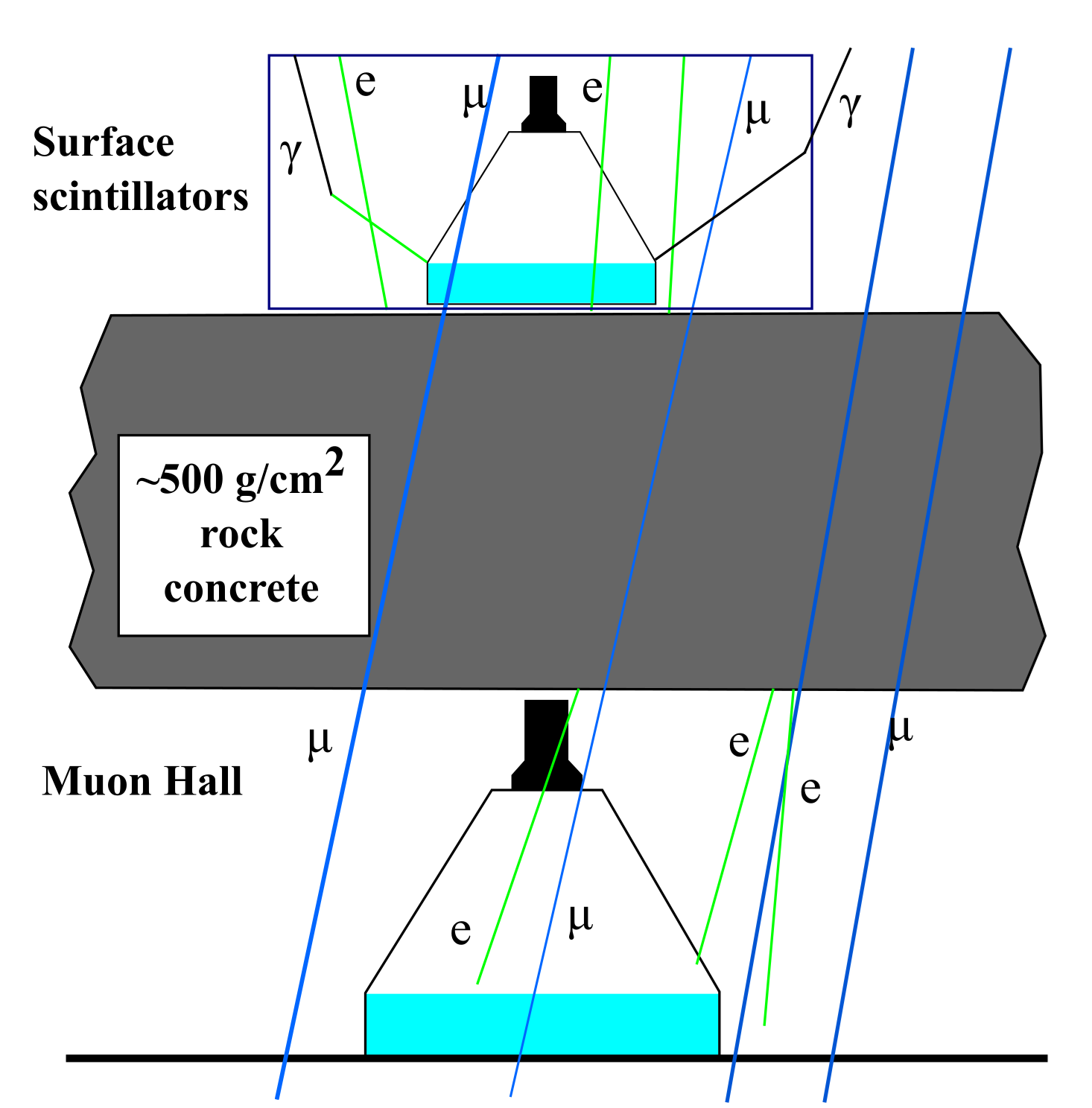}
	\caption{ Scheme of the muon telescope used in the Small Cherenkov array. The telescope is two scintillation detectors mounted on top of each other and separated by a concrete.
		\label{ykt-fig-telescope}}
\end{figure}

As part of the Small Cherenkov Array, there are three muon telescopes with $\varepsilon_{thr.} \geq$ 1 GeV. The scheme of the telescope is shown in Fig. \ref{ykt-fig-telescope}. The muon telescope consists of two scintillation detectors, ground and underground, located one above the other with 1 and 2 m$^2$ area respectively. 

\begin{figure}
	\begin{minipage}[h]{0.4\linewidth}
		\center{\includegraphics[width=1.0\linewidth]{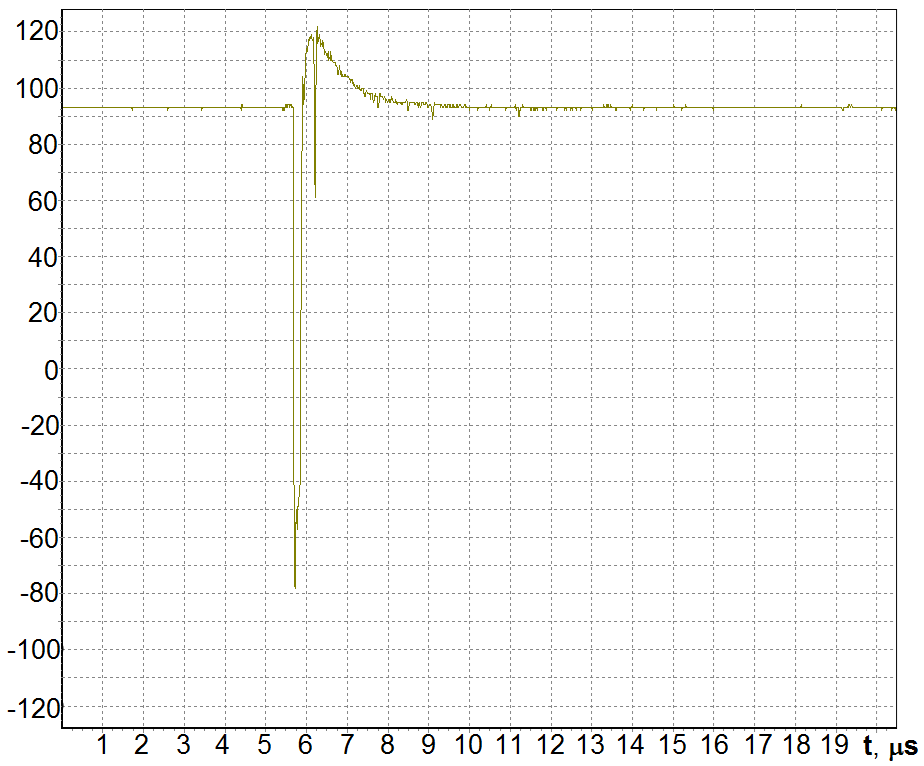} \\ a)}
	\end{minipage}
	\begin{minipage}[h]{0.4\linewidth}
		\center{\includegraphics[width=1.0\linewidth]{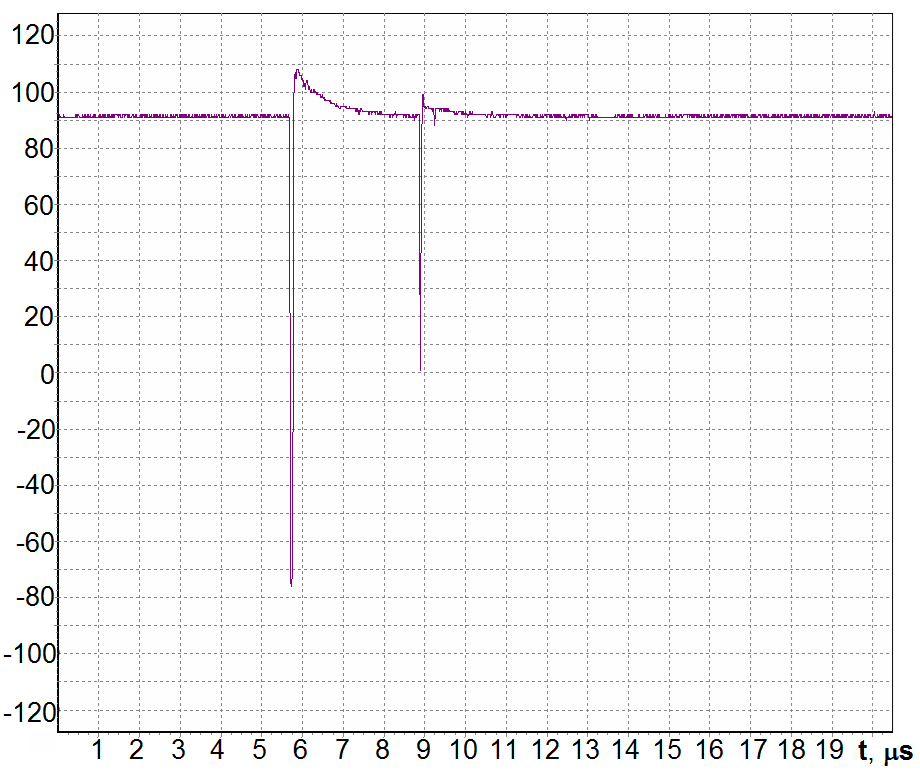} \\ b)}
	\end{minipage}
	\caption{ Air shower event 05.01.18, T = 00:12:55. a) ground detector with s = 1 m$^{2}$. b) underground detector with $\varepsilon_{thr.}\geq$ 1 GeV \label{ykt-fig-signal_sweep}}	
\end{figure}

Muon telescopes, in addition, record the spatio-temporal image of signals \cite{Knurenko2013409, Knurenko201781}. Examples of registration of pulses from the passage of charged particles and muons in an individual shower are shown in Fig. \ref{ykt-fig-signal_sweep}.

In Fig. \ref{ykt-fig-signal_sweep}a, the signal sweep consists of several peaks separated by the time of the particles arrival at the detector, and in aggregate corresponds to the total signal from electrons, gamma rays, and muons. The concrete cuts off the electromagnetic component of the shower, and the underground scintillation detector measures only the response from muons. Muon component has narrower signal with single-pulse structure (Fig. \ref{ykt-fig-signal_sweep}b). 
Comparing both figures we can say that the contribution of electrons and gamma rays to the signal is small, because the signal of the underground detector is very close in amplitude to the surface detector signal. At the zenith angle $\theta\geq$45$^\circ$, the air shower disk at the ground level mainly contains muons, which are distributed in time from 0 to 3 $\mu s$ \cite{Knurenko2013409, Knurenko201781}.

\begin{figure}
	\includegraphics[width=0.7\textwidth]{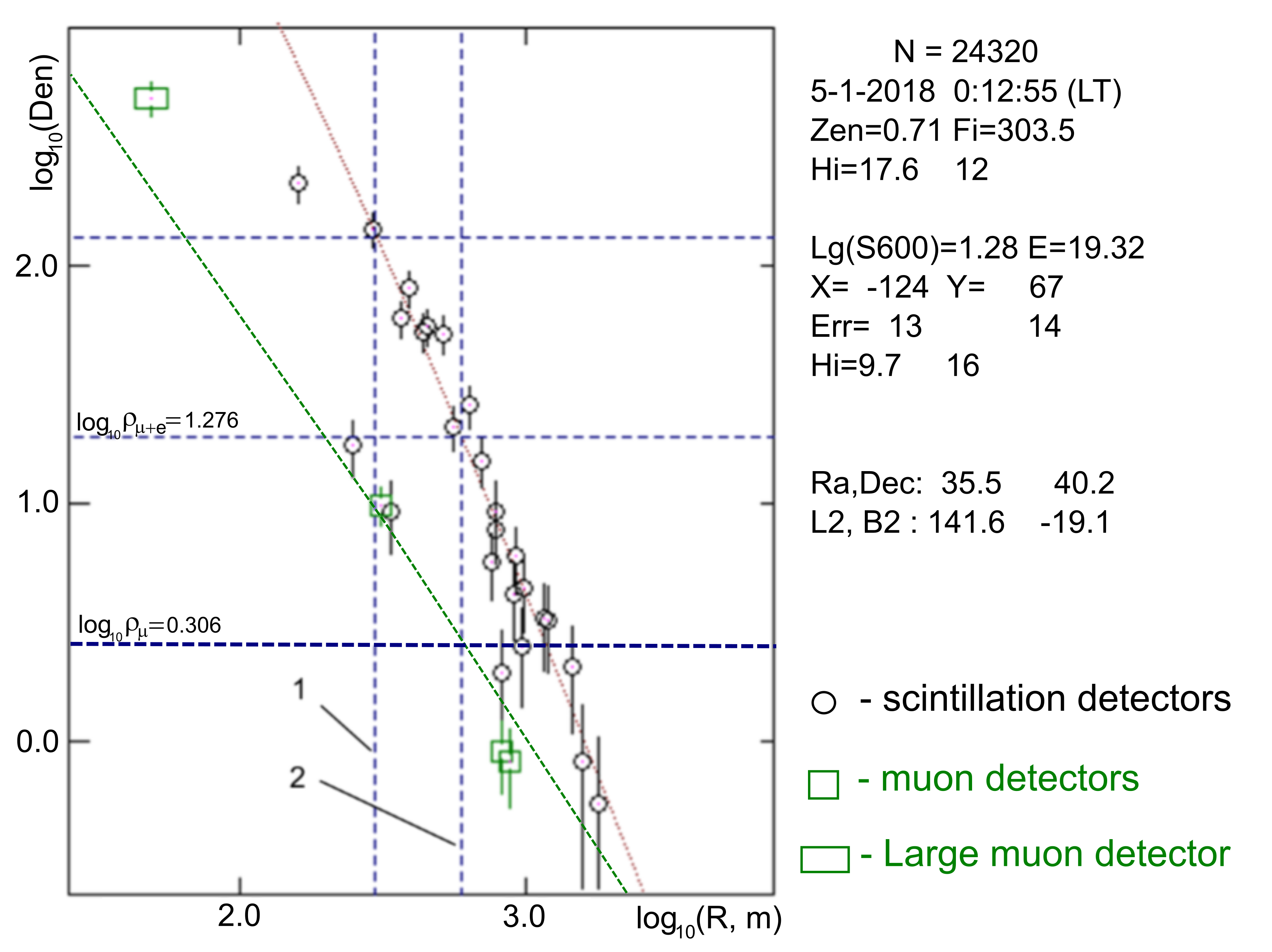}
	\caption{Lateral distribution function of charged particles flux density (circles) and muons (squares) in shower with energy $E_{0}$ = 20.9 EeV registered at 05.01.18. OX axis represents logarithm base 10 of distance R; OY axis represents logarithm base 10 of charged particles flux density. 1 --- $\rho_{\mu+e}$ (300) --- flux density of charged particles and muons at the distance of 300 m from the air shower; 2 --- $\rho_{\mu+e}$(600) --- flux density of charged particles and muons at the distance of 600 m from the air shower. The value of $\log_{10}{\rho_{\mu}}$ = 0.306 for this shower and parameter $\rho_{\mu}/\rho_{\mu+e}$ = 0.107. Zenith angle $\theta$ = 44.8$^{\circ}$. \label{ykt-fig-ldf}
	}
\end{figure}

Fig. \ref{ykt-fig-ldf} shows lateral distributions of charged particles (circles) and muons (squares) of an individual shower with energy $E_{0}$ = 20.9 EeV and zenith angle $\theta$ = 44.8$^{\circ}$. Muon LDF is measured relative to the air shower axis and given as a function of muon density divided by detector area $\rho_{\mu}(R)$ = n/s$\cdot \cos{\theta}$, where n --- number of muons and s --- area of the detector.

Air shower event from 5.01.2018 is not an ordinary one, because of its inclination ($\theta$ = 44.8 $^\circ$) and low measured muons compared to total charged particles. It may happen in the case of $X_{max}$ of air shower is at or below the sea level. In this case, it is initial stage of air shower development in the atmosphere and formula (\ref{ykt_eq_5}) for energy estimation is not applicable. It’s better to use formula (\ref{ykt_mu_e}), since electron-photon component of the shower is at the maximum of development and its energy almost equal to air shower energy. It is possible that this shower has been produced by primary gamma ray of ultra-high energy with the depth of maximum near the sea level. Depth of maximum $X_{max}$ estimation results in 953$\pm$55 g$\cdot$cm$^{-2}$ \cite{Knurenko2018107}.

The dotted line (Fig. \ref{ykt-fig-ldf}) shows the data approximation for charged particles using function (\ref{ykt_eq_1}). Dashed approximation line is for muons. The dashed horizontal lines indicate the flux density of charged particles and muons for distances R = 300 and 600 m from the shower axis, which are used as classification parameters at the Yakutsk experiment. The classification parameter for showers with energies below 0.5 EeV, we use $\rho_{\mu+e}$ (R = 300), and for energies greater than 0.5 EeV we use $\rho_{\mu+e}$ (R=600). In this work we use only $\rho_{\mu+e}$ (R=600). This parameter correlates with air shower energy and is used for the fast estimation of shower energy and the selection of showers for a particular physical tests.

\subsection{Muon flux density $\rho_{\mu}$ at the distance of 600 m from the air shower axis.}

The values of the muon flux density, measured at different distances from the axis in individual showers, are used to study the properties of muons in air showers of ultrahigh energies. In particular, at the first stage of measurements of the muon component, the average LDF of muon flux was obtained \cite{Glushkov1983}. The inclination of the muon LDF was significantly flatter (b $\sim$ 2) than the inclination of the charged particles LDF (b $\sim$ 4); therefore, the pole of the function was taken as unity (b $\sim$ 1). In our case --- to analytically describe the form of the muon LDF --- we used a function similar to the Greisen function, but with coefficients describing a shape of the experimental muon LDF. The parameters of function (\ref{ykt_eq_3}) were selected by the maximum likelihood method. 
\begin{equation}\label{ykt_eq_3}
\rho_{\mu}(R) = 
\frac{N_{\mu}}
{2\pi R^{2}_{0}}
\cdot\frac{b-2}
{x}
\cdot\frac{1}
{(1+x)(b-1)}
\end{equation}
where x = $\frac{R}{R_{0}}$,  $N_{\mu}$ --- total number of  muons, b --- slope parameter of the muon LDF, and $R_{0}$ --- parameter that depends on atmospheric model (in our case it's 70).

The function (\ref{ykt_eq_3}) well describes average muon LDF in the range of 100-800 m distances and air showers with energy $E_0 \geq$ 0.5 EeV with zenith angles $\theta \leq$ 60 $^\circ$. Equations (\ref{ykt_eq_1}) and (\ref{ykt_eq_3}) were used to calculate parameters $\rho_{\mu+e}$ and $\rho_{\mu}$.

\subsection{Dependence of muon fraction parameter $\rho_{\mu}/\rho_{\mu+e}$ zenith angle and flux density of charged particles $\rho_{\mu+e}$}

The dependence of the parameter $\rho_{\mu}/\rho_{\mu+e}$ on zenith angle and parameter $\rho_{\mu+e}$ was obtained from 1995-2015 data \cite{Knurenko201175}. At the Yakutsk array, the parameter $\rho_{\mu+e}$ was adopted as a classification parameter, since fluctuations of charged particles are small at this distance and the density of charged particles at R = 600 m is proportional to the shower energy \cite{Ivanov2007104, Knurenko200683}.

\begin{figure}
	\includegraphics[width=0.8\textwidth]{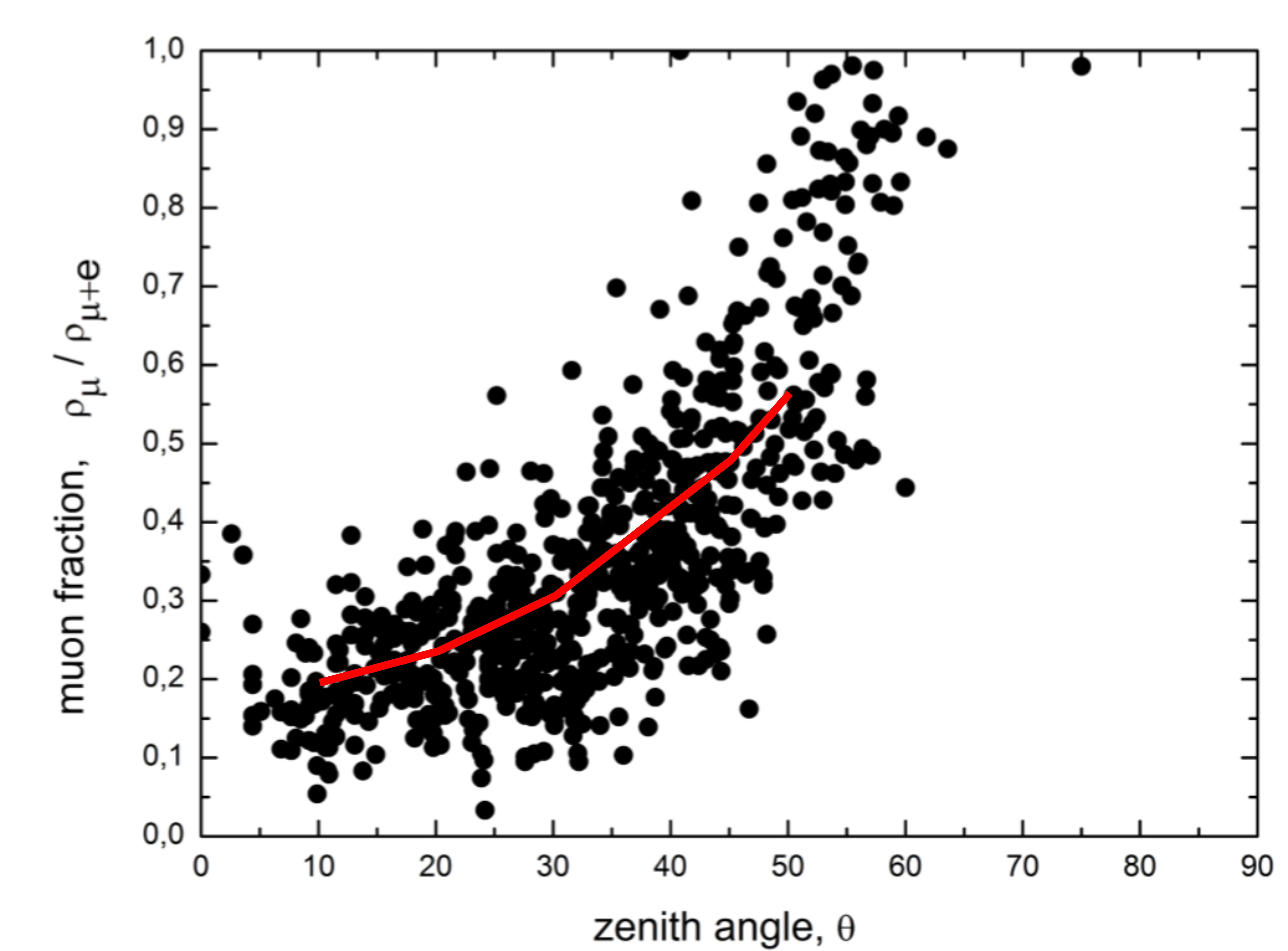}
	\caption{Dependence of muon fraction on zenith angle. Dots represents individual showers. OX axis --- zenith angle [$^\circ$]; OY axis --- muon fraction. \label{ykt-fig-muon_zenith}
	}
\end{figure}

Fig. \ref{ykt-fig-muon_zenith} shows the dependence of the ratio $\rho_{\mu}/\rho_{\mu+e}$ on the zenith angle. The data consist of air showers with different energies and with zenith angles $\theta \leq$ 60 $^\circ$. The number of muons sharply increases at large zenith angles, i.e. the shower almost entirely consists of muons with a small portion of electrons. Significant scatter of dots is due to uncertainty in zenith angle determination at the Yakutsk experiment and due to mass composition of particles producing air showers. It is known that showers with high muon content were produced by heavier nuclei, and showers with low muon content were produced by lighter nuclei --- protons, helium nuclei, or even ultra-high energy gamma rays. In general, we can assume that showers at the lower boundary of the distribution (Fig. \ref{ykt-fig-muon_zenith}) are produced by light nuclei and showers at the upper boundary of the distribution are produced by heavier nuclei. 

From the detailed analysis of the data we derived dependence of the relationship $A = \rho_{\mu}/\rho_{\mu+e}$ with the zenith angle $\theta$ and the classification parameter $\rho_{\mu+e}$ (eq. \ref{ykt_eq_4}). It was used to normalize parameter $A_{\theta}$ to the vertical angle.

\begin{equation}\label{ykt_eq_4}
A_{\theta=0^{\circ}}=
A_{\theta}
-(1.98\pm 0.45)
\cdot\log_{10}{\sec{\theta}}
+(0.06\pm 0.02)
\log_{10}{\rho_{\mu+e}}
\end{equation}

\subsection{Flux density of muons parameter dependence on air shower energy}

It is known that secondary particles lose most of its energy on air ionization (mainly electrons and positrons production), and the rest of energy is lost on the processes of splitting and interaction of the shower hadron component with air nuclei and is carried by high-energy muons and neutrinos to sea level. 

At the Yakutsk experiment, the shower energy is determined empirically using the integral characteristics of the main components of the air shower: the total number of charged particles $N_{s}$, the total number of muons $N_{\mu}$, and the total flux of Cherenkov light $\Phi$. This is the energy balance method for all particles in air shower \cite{Ivanov2007104, Knurenko200683}. The method suggests that the total shower energy can be determined as a sum of ionization losses of each air shower components. It is based on measurements of the total flux $\Phi$ of Cherenkov light, which allows determination of energy transferred to electron-photon component of the shower, $E_{em}$ = k(x, $P_\lambda$)$\cdot \Phi$, where k(x, $P_\lambda$) --- approximation coefficient (calculated value), takes into
account the transparency of the real atmosphere, the nature of the longitudinal development of the shower (energy spectrum of secondary particles and its dependence on the age of the shower) and expressed through the depth of the maximum of air shower $X_{max}$, measured at the array.

The full flux of Cherenkov light was taken as the basis of the energy balance method, since it reflects the ionization losses of the electron-photon component of the air shower. From Table \ref{ykt_tab_1} it follows that the observation time are different for different components. All components measured simultaneously only when atmospheric conditions are favorable for optical observations, since charged component detectors do not depend on atmospheric conditions. Air showers with measured Cherenkov light energy can be determined by formula (\ref{ykt_q400}), which is derived from correlation between energy $E_0$ and Cherenkov light flux density Q(400) at 400 m distance. 
\begin{equation}\label{ykt_q400}
\log_{10}{E_0}=
17.89 + 
1.03\cdot
(\log_{10}{(Q(400))}-7)
\end{equation}
Since observation time of charged component is the longest at the Yakutsk experiment air shower energy is determined by correlation of $E_0$ and $\rho_{\mu+e}$ using formula:
\begin{equation}\label{ykt_mu_e}
\log_{10}{E_0}=
17.68+
\log_{10}{(\rho_{\mu+e}(R=600,\theta=0^\circ))}
\end{equation}
In this case the parameter $\rho_{\mu+e}$ was transformed to the vertical with absorption path equal to $\lambda\sim$ 500 g$\cdot$ cm$^{-2}$\cite{Glushkov199357}:
\begin{equation}\label{ykt_zenith_transform}
\rho_{\mu+e}(R=600,\theta=0^\circ) = 
\rho_{\mu+e}(R=600,\theta)\cdot
exp\left(X_{0}
\frac{(\sec{\theta}-1)}{\lambda}
\right)
\end{equation}
Here, the zenith-angular dependence of the parameter $\rho_{\mu+e}$ was found by analyzing showers in a wide range of zenith angles, and the path $\lambda$ was determined by the method of lines of equal intensities.

The shower energy can also be determined by the muon component, using the correlation between $E_0$ and $\rho_{\mu+e}$ and the range for muon absorption path equal to $\lambda\approx$  1900 g$\cdot$cm$^{-2}$ (eq. (\ref{ykt_eq_5})) \cite{Glushkov199357}. It should be noted that formula (\ref{ykt_eq_5}) is applicable only for air showers with a cascade curve maximum in the atmosphere equal to $X_{max}\leq$ 900 g$\cdot$cm$^{-2}$, i.e. for fully developed showers. 

Within framework of formulas (\ref{ykt_q400}), (\ref{ykt_mu_e}) and (\ref{ykt_eq_5}) air shower energy estimation is in agreement within 10$\%$ accuracy. Systematic uncertainty of energy balance method \cite{Ivanov2007104, Knurenko200683} is 25$\%$. 

\begin{figure}
	\includegraphics[width=0.8\textwidth]{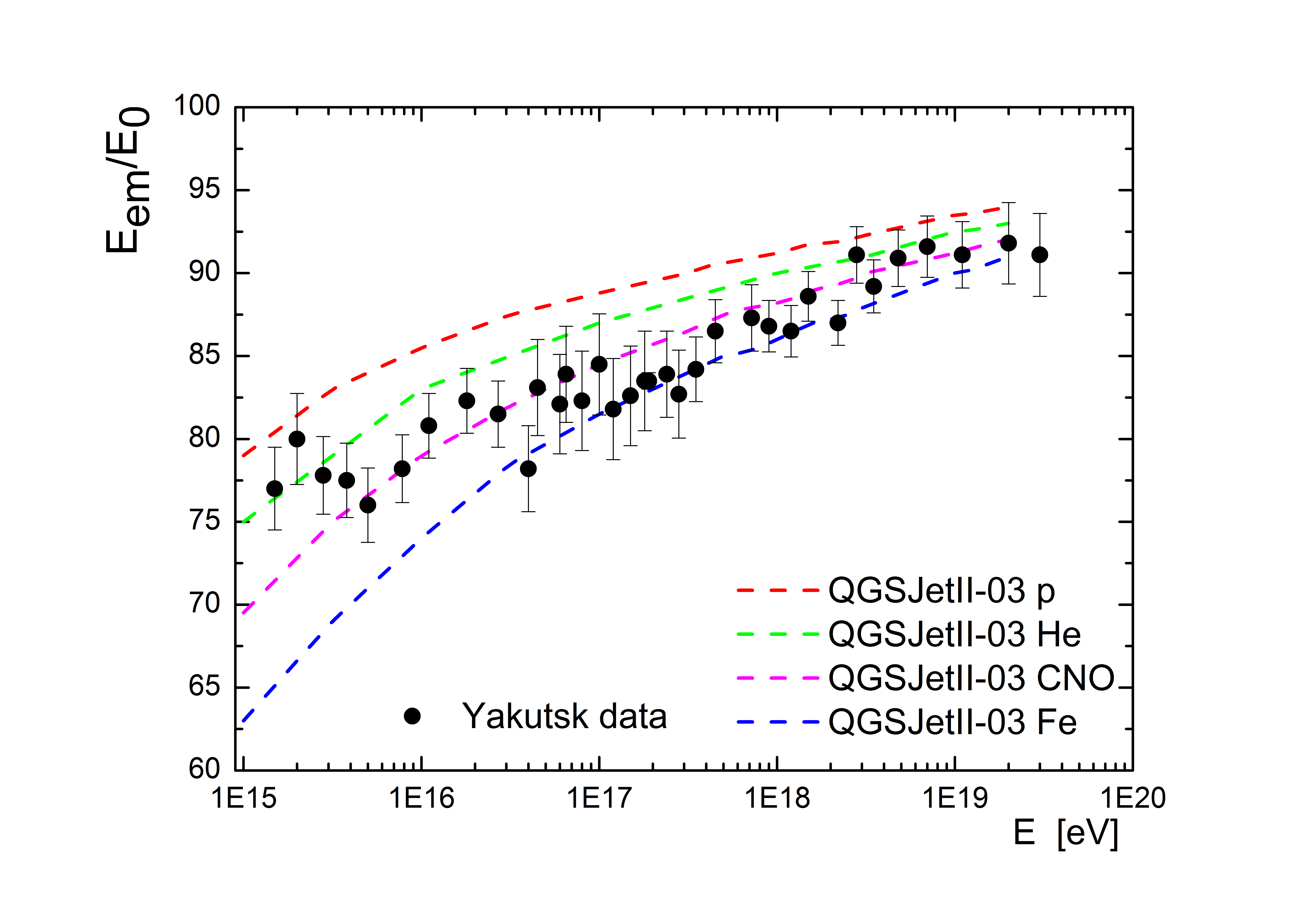}
	\caption{ Fraction of energy transferred to electromagnetic component by Cherenkov light data at the Yakutsk array and hadronic interaction model QGSJetII-03 for proton p, helium He, CNO nuclei and iron Fe. \label{ykt-fig-longitudinal}
	}
\end{figure}

The result obtained at the Yakutsk experiment on the fraction of energy used for ionization of air is shown in Fig. \ref{ykt-fig-longitudinal}. QGSJetII-03 simulations for different primary nuclei are also given there. As can be seen from the figure, the electromagnetic component accounts for 75-90$\%$ of the total shower energy in the range 1 PeV to 10 EeV; the same picture is observed in the simulations of hadron interaction models. Fraction of scattered energy depends on the atomic weight of the primary particle that formed the air shower, as discussed in \cite{Knurenko200520}. Therefore, the empirical estimation of shower energy by the energy balance method is closer to the real case, compared to energy estimation by using hadronic interaction models.

\begin{figure}
	\includegraphics[width=0.8\textwidth]{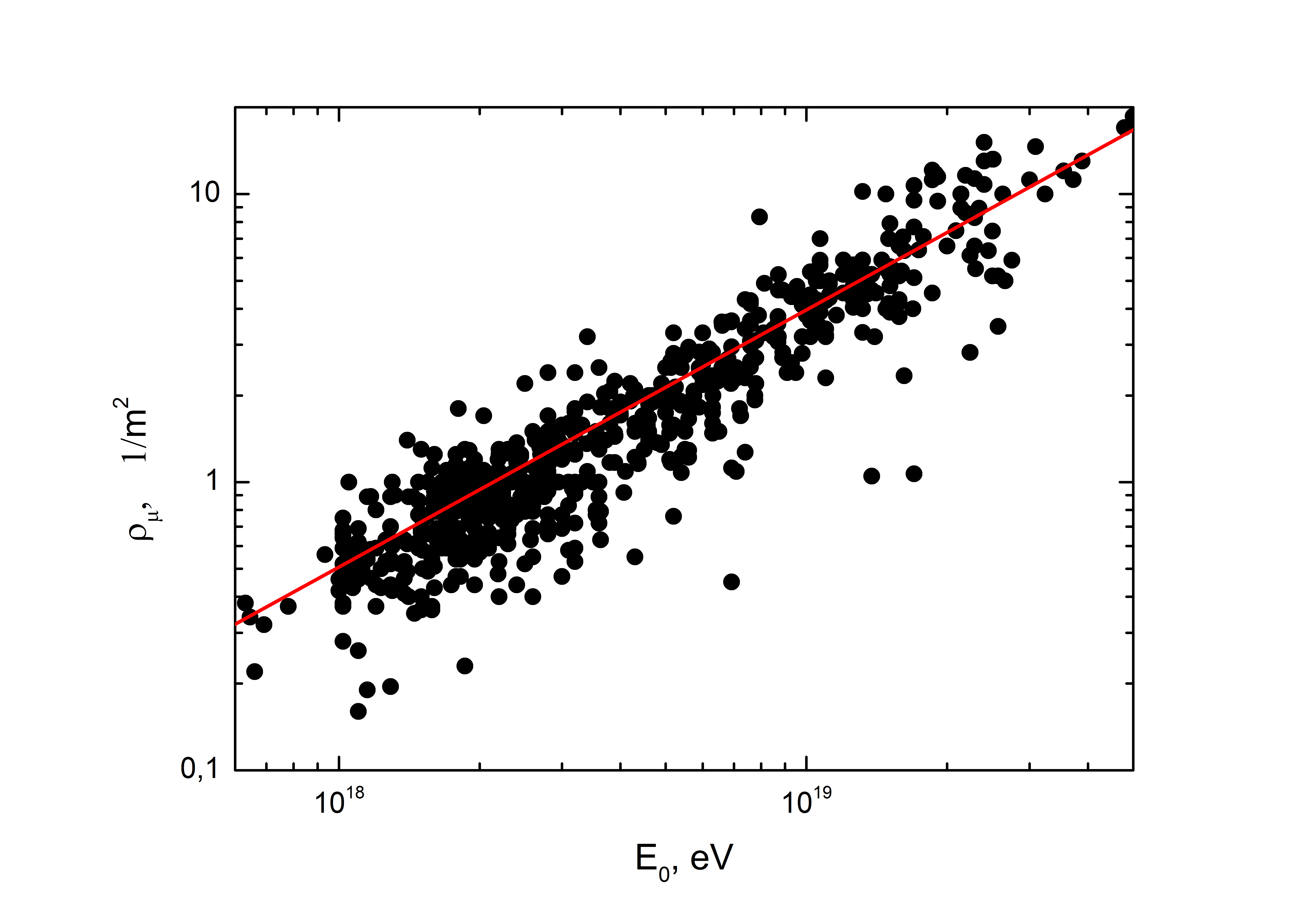}
	\caption{ Correlation of classification parameter $\rho_{\mu}$ and air shower energy $ E_{0} $. Here, $E_{0}$ is determined by method of balancing all components energies of air showers, and $\rho_{\mu}$ determined by formula (\ref{ykt_eq_3}) with no conversion to vertical. \label{ykt-fig-muon_energy}
	}
\end{figure}

According to simulations \cite{KnurenkoDis, IvanovDis}, flux density of charged particles $\rho_{\mu+e}$ and flux density of muons $\rho_{\mu}$ are proportional to shower energy and have small shower-to-shower fluctuations, and it makes them great classification parameters of air shower events. The relationship between the energy $E_{0}$ and $\rho_{\mu}$ was found based on the correlation of the energy and density of muons at a distance of 600 m from the shower axis (Fig. \ref{ykt-fig-muon_energy}). Here, the shower energy was determined by the energy balance method, and $\rho_{\mu}$ from the LDF of muons using formula (\ref{ykt_eq_3}).
\begin{equation}\label{ykt_eq_5}
\log_{10}{E_{0}} = 
18.33 + 1.12
\cdot \log_{10}{\rho_{\mu}(R=600,\theta)}                                                    
\end{equation}
Eq. (\ref{ykt_eq_5}) was further used for energy $E_{0}$ estimation in individual showers by $\rho_{\mu}$ parameter. 

We, also, determined ratio of $\rho_{\mu}/\rho_{\mu+e}$ for 600 m distance. This ratio, for fixed energy, is the most sensitive characteristics for the mass composition of cosmic rays. 

\subsection{Selection of air showers with muons}

For this work, we used data obtained for 20 years (1995-2005) of continuous operation at the Yakutsk experiment. This period of time air shower measurements were conducted by muon detector with large area, six muon underground detectors with medium area, three muon telescopes with smaller area (see Section \ref{sec_muon}), ground-based scintillation detectors and Cherenkov light detectors (Fig. \ref{ykt-mc-experiment}). The selection criteria for this work were: the axis of the showers should be within a circle with a radius of 1.2 km from the center of the array; the muon measurements at a distance of 400-800 m, which allows determination of the muon flux density at 600 m distance with good accuracy including the periphery of the muon radial distribution; the energy of the showers should be greater than 5 EeV; zenith angles of the showers $\theta \leq$ 60$^\circ$. Out of 1365251 events, 802 showers were selected with a measurement accuracy of 5-10$\%$. Twenty-five percent of them were showers with energies greater than 10 EeV.

\begin{figure}
	\includegraphics[width=0.5\textwidth]{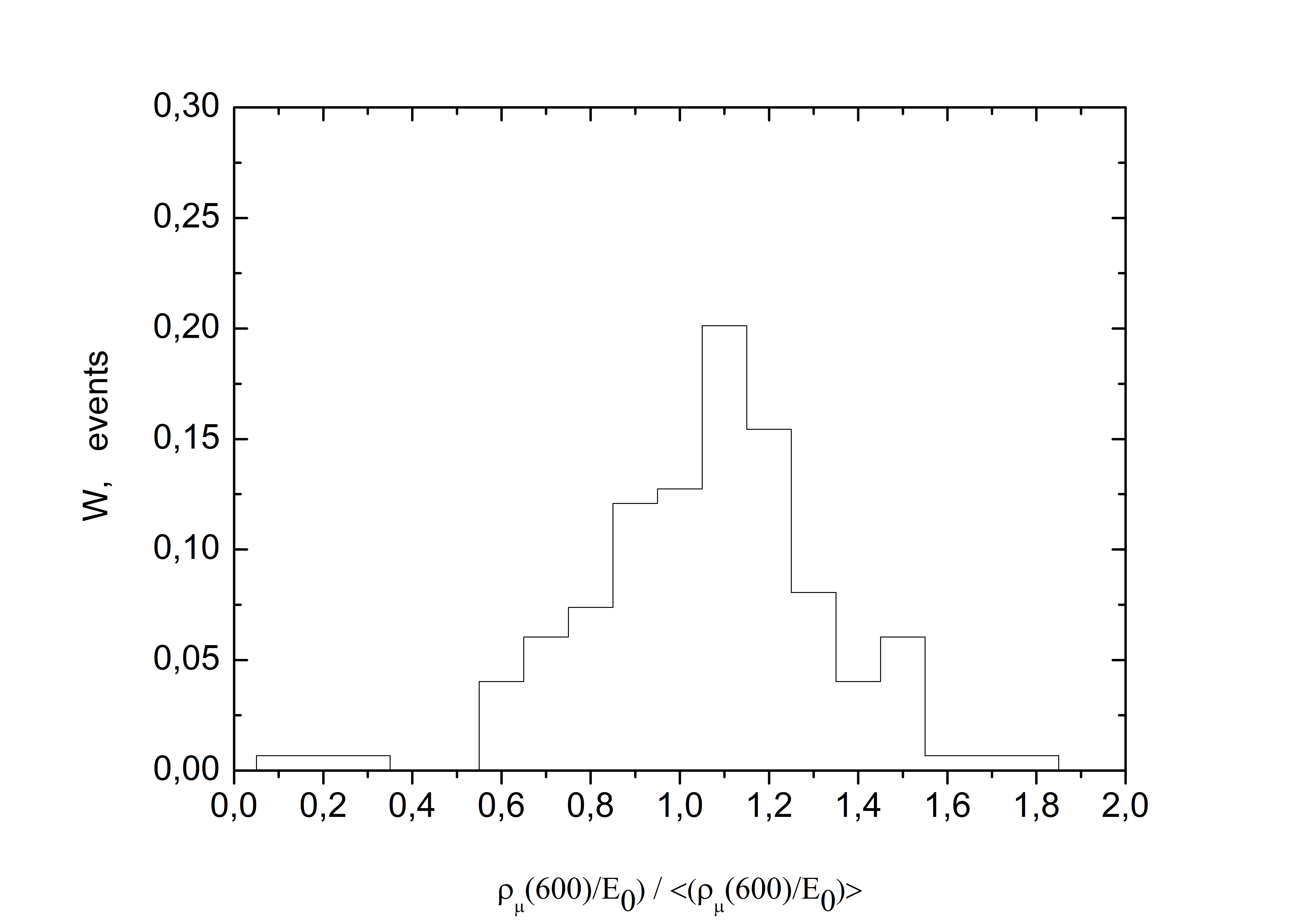}
	\caption{Distribution of muon numbers in showers with energy 5-50 EeV and zenith angles $\theta \leq$ 60 $^{\circ}$ by the Yakutsk data. Data are normalized to vertical ($\theta$ = 0$^{\circ}$), according to dependence of muon fraction $\rho_{\mu}/\rho_{\mu+e}$ on zenith angle $\theta$ by the Yakutsk experiment. \label{ykt-fig-muon_distr}
	}
\end{figure}

Distribution of showers selected by the parameter $\rho_{\mu}/\rho_{\mu+e}$ after conversion to the vertical $\theta$ = 0 $^\circ$ shown in Fig. \ref{ykt-fig-muon_distr}. It is shown that the experimental distribution has several unexpressed maxima, which, according to the QGSJetII-04 simulations, are produced by primary particles with different masses \cite{Ostapchenko201183, Kalmykov199356, Ostapchenko2006151}.

\begin{figure}
	\includegraphics[width=0.5\textwidth]{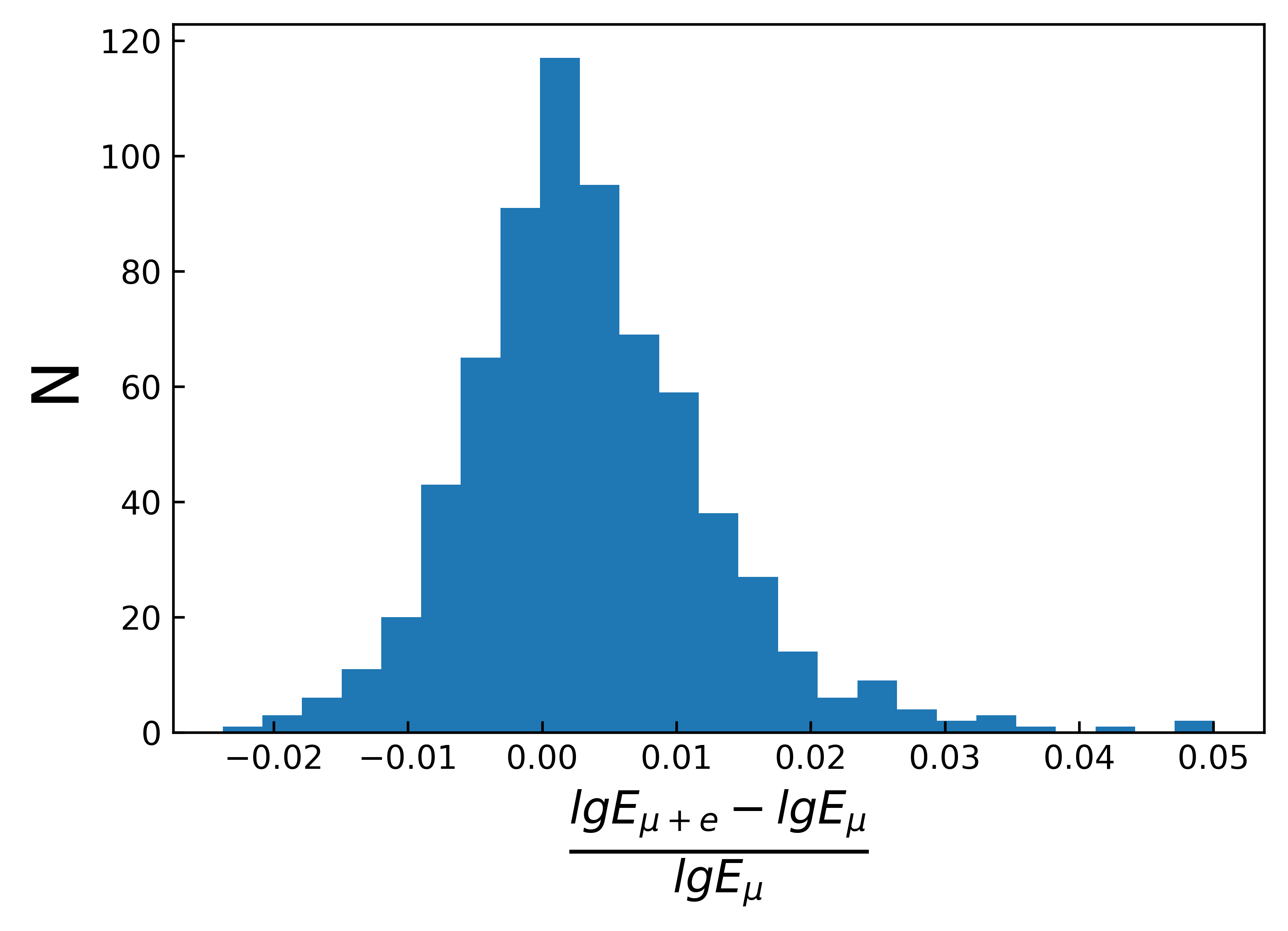}
	\caption{ Comparison between energy estimated by muon component and charged component \label{ykt-fig-comparison}
	}
\end{figure}

Air shower energies were estimated by muon component (eq. \ref{ykt_eq_5}) and comparison with energy estimated by charged component is shown on Fig. \ref{ykt-fig-comparison}

In Fig.\ref{ykt-fig-comparison}, $E_{\mu+e}$ --- the energy estimated by charged component, determined by eq. \ref{ykt_mu_e}, $E_\mu$ --- the energy estimated by the muon component, determined by eq. \ref{ykt_eq_5}. As can be seen, estimations of the air showers energies from the muon and charged components are in a good agreement. Therefore, the selection of showers by muons should not affect this analysis.

\section{Mass composition of cosmic rays with highest energies}
\subsection{Experimental data comparison with simulations}

\begin{figure}
	\includegraphics[width=0.7\textwidth]{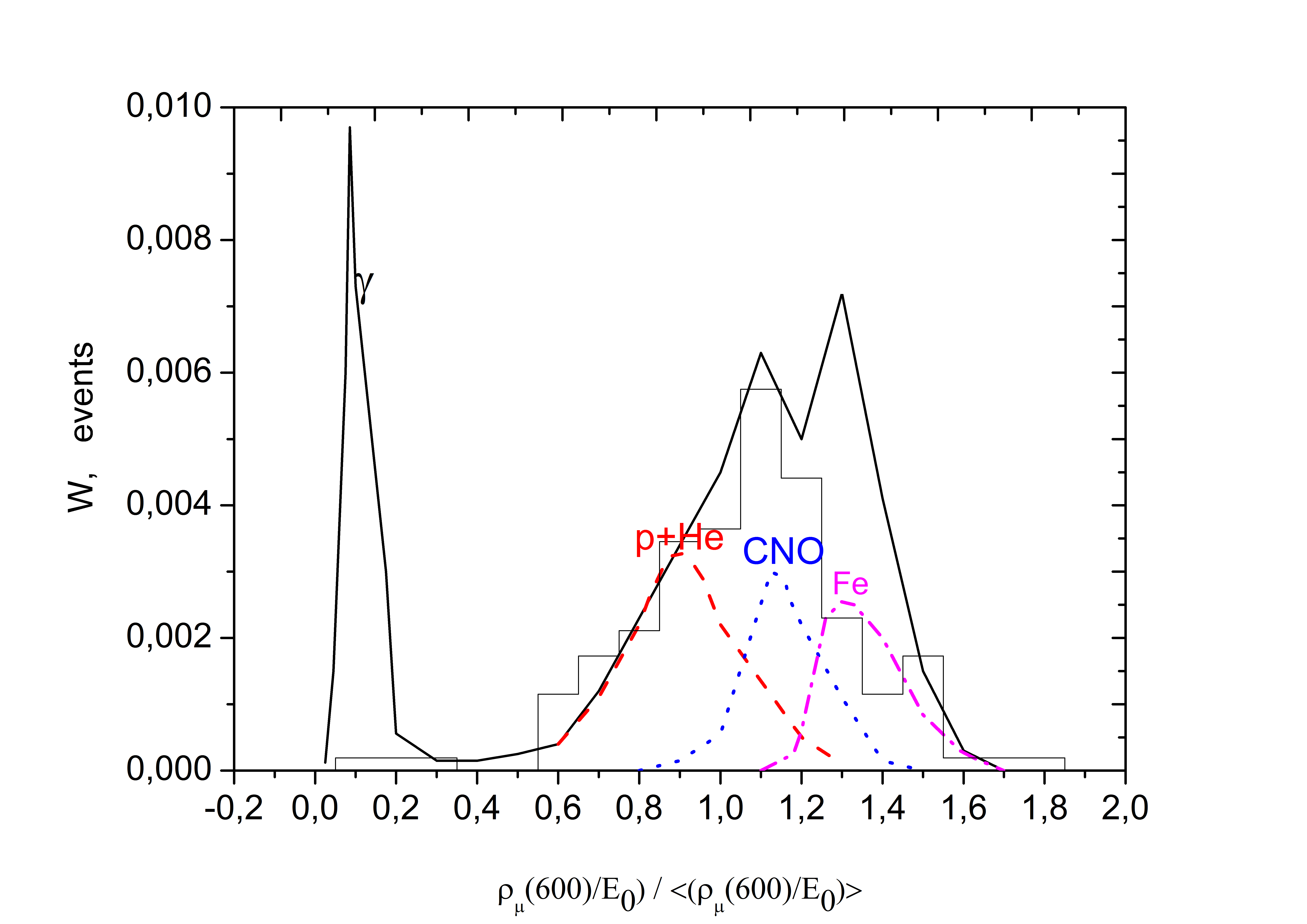}
	\caption{ Simulations of muon fraction in air showers according to QGSJETII-04 for different primaries (curves). Histogram is experimental data with energy 5-50 EeV and zenith angles $\theta \leq$ 60$^{\circ}$. Curves denote: dashed curve --- p + He, dots --- CNO, dash-dotted curve --- Fe. The solid curve indicates the total value of p + He + CNO + Fe nuclei. \label{ykt-fig-muon_distr_sim}
	}
\end{figure}

For a quantitative estimation of the atomic weight of a particle producing a shower, we used the calculations based on the QGSJETII-04 and the experimental data shown in Fig. \ref{ykt-fig-muon_distr}. The experimental data were compared with the calculations performed for the primary gamma ray, proton, carbon and iron nuclei. The calculations took into account the response of underground and ground scintillation detectors to muons with a threshold $\varepsilon_{thr.} \geq$ 1 GeV provided that more than one muon passes through the detectors. The calculations were normalized to the general statistics of the selected showers, so they could be directly compared with experimental data. Fig. \ref{ykt-fig-muon_distr_sim} shows such a comparison. From Fig. \ref{ykt-fig-muon_distr_sim} it can be seen that each of the calculated components ($\gamma$, p+He, CNO, Fe) has its own boundaries. They are formed due to fluctuations in the development of air showers and instrumental errors in the measurement of muons. For example, particles have the following boundaries: gamma ray $\gamma$ --- (0-0.3), proton and helium p+He --- (0.4-1.3), CNO nuclei --- (0.8-1.5) and iron Fe --- (1.1-1.8). With good accuracy in measuring muons and charged particles by relative content, it is possible to divide showers into those produced by a gamma ray, proton, or iron nucleus.

It can be qualitatively said that the composition of cosmic rays in the energy range 5-50 EeV is represented by a mix of nuclei with a fraction of protons, helium and carbon nuclei (70-80)$\%$, the fraction of heavy nuclei is less than (20-30)$ \% $, and (1-2)$ \% $ in the sample are candidates for primary gamma rays

\subsection{Discussion}

Comparing the calculation results with the experimental data, which are presented in Fig. 8, it should be noted that the distribution for each particle is localized at certain boundaries and when compared with the experiment, creates the effect of some similarity in the form of the shown distributions. For example, in the left part of Fig. \ref{ykt-fig-muon_distr_sim}, we can distinguish a group of showers with a low muon content interval (0-0.3), a separate peak distinguishes a group of showers in the interval (0.4-1.3), a group in the interval (0.8-1.5) and the last group (1.1-1.8). Based on the calculations, we can conclude: the first group of showers is produced by primary gamma rays; the second by protons and helium nuclei --- air shower events with a relatively low muon content; the third group by CNO nuclei; the fourth by iron nuclei. The number of showers supposedly produced by protons and helium nuclei are (40-50) $ \% $. Showers that are possibly produced by primary gamma rays are (1-2) $ \% $. The remaining showers are formed by nuclei of heavier chemical elements: CNO and iron Fe nuclei. This result is preliminary because of the small statistics, high experimental uncertainty in the measurement of muons and method of air shower processing, which results in the “smearing” of the histogram in Fig. \ref{ykt-fig-muon_distr_sim}. To improve the quality of comparison, the accuracy of measuring muons and charged particles at the level of (3-5) $ \% $ is required \cite{Ostapchenko201183}, which is not yet comparable with the current uncertainty of the Yakutsk experiment 5-15$ \% $. 

It should be noted that most of the selected showers are grouped on the left side of the figure (Fig. \ref{ykt-fig-muon_distr_sim}), i.e. contain a small amount of muons, which indicates light component. Taking into account the accuracy of the experiment and the calculations using the QGSJETII-04 model, we can conclude that the bulk of air showers with energies of 5-10 EeV are produced by protons p and helium nuclei He. A small number of showers is produced by heavier nuclei and gamma rays $\gamma$. The result obtained in this work on the composition of cosmic rays of highest energies does not contradict the results from \cite{Berezhko201231,Hanlon201819,Aab201490,Abbasi201564,Abbasi201999,Bellido2018506}. It indicates the mass composition of cosmic particles in the energy region of 5-10 EeV is represented by a mixture: with a high content of protons p, light nuclei like He, an insignificant content of gamma rays $\gamma$ and iron nuclei Fe.

\section{Conclusion}

The muon component measured at the Yakutsk array made it possible to obtain information on the lateral distribution, relative content, and spatio-temporal distribution of muons in the air shower disk with energies of 5-50 EeV. Thanks to the long-term (over 40 years) measurements of muons at the array, it was possible to create a database for further study of the characteristics of muons in showers of highest energies. It turned out that the lateral distribution of muons is more flatter compared to the lateral distribution of charged particles (Fig. \ref{ykt-fig-ldf}). Such feature of the radial development of muon component and electron-photon component of the shower at sea level, is associated with the nature of the primary particle interaction with the nuclei of air atoms and with the longitudinal development of air shower --- the depth of shower development maximum $X_{max}$\cite{Knurenko199846, Egorov2018301, Knurenko2017230}.

It was used in the present work for an independent estimation of the mass composition of cosmic rays in the region of highest energies. For this purpose, the parameter $\rho_{\mu}/\rho_{\mu+e}$ was chosen, which was measured with good accuracy on the array and is sensitive to the mass of the primary particle producing the air shower. From the measurements, the dependence $\rho_{\mu}/\rho_{\mu+e}$ on the zenith angle and shower energy was established. The dependence made possible to normalize all selected showers to the same conditions for the development of muons in the atmosphere and use these data to estimate the mass composition of cosmic rays. 

Fig. \ref{ykt-fig-muon_distr_sim} shows the experimental energy-normalized distribution of the parameter $\rho_{\mu}/\rho_{\mu+e}$ as a function of the number of muons in the average shower at a fixed energy. QGSJETII-04 simulations for different primary particles are also plotted there. It can be seen from Fig. \ref{ykt-fig-muon_distr_sim} that the distribution has several maxima, and these maxima relate to particles with different atomic weights. A quantitative comparison of the experimental data with the calculations showed: a) the group of showers with energies of 5-10 EeV is presumably produced by protons p and He helium nuclei, which is 40-50$ \% $ of the total number of showers in the sample; b) (1-2)$ \% $ of showers are possibly produced by primary gamma rays $\gamma$; c) other showers are produced by nuclei of heavier chemical elements, e.g. CNO and iron nuclei Fe.

\section* {Acknowledgments}
We wish to thank all of the staff of the Yakutsk Array for the opportunity to use experimental data and useful discussion during the article preparation.

\section*{References}

\bibliography{mybibfile}

\end{document}